# How can AI accelerate the green transition?

Jacques Sainte-Marie[1] , Marie-Fleur Simmet[2] and Fanny Terrier[3]

In his essay The Gift, published in 1925, the sociologist Marcel Mauss emphasised the 'total' scope of the phenomenon of gift- and counter-gift-giving, noting that it *'expresses, simultaneously and all at once, all manner of institutions: religious, legal and moral (...) ; economic (...) ; not to mention the aesthetic phenomena to which these facts give rise (...)*." The emergence of artificial intelligence (AI), which is suddenly and profoundly disrupting individual and collective lives, bears a resemblance to the "total social fact" described a hundred years ago. AI refers to technologies designed to simulate human intelligence, namely an *"automated system that (...) deduces, from received inputs, how to generate output results such as predictions, content, recommendations or decisions that may influence physical or virtual environments*." [4] (OECD).

Having become central to human relations, AI relies on physical and globalised infrastructure and therefore, also has an impact on the environment. This paper analyses some of these effects and how they are perceived by society.

This text offers an exploration of the links between AI and the ecological transition, seeking to identify new opportunities and constraints. To this end, it will identify the barriers to the ecological transition (1) and the environmental impacts attributable to AI (2). It will then go on to describe how to live in a constrained world (3) while AI accelerates scientific discovery (4) and profoundly transforms society (5).

## 1. Some barriers to the ecological transition

### The limits of individual and collective action in the face of climate change

The consequences of ongoing climate change are becoming increasingly frequent and severe (heatwaves, droughts, floods, etc.), affecting nature, ecosystems and our lives.

As the role of human activities in climate change is now well established, we are both culprits and victims. The frequency and intensity of these events are prompting some members of the public to seek collective strategies or individual practices aimed at minimising their environmental footprint. This approach faces certain limitations:

- France accounts for less than 1% of the world's population, which dilutes the impact of actions taken individually or by a small group.
- For a French person, carbon neutrality requires reducing their environmental footprint from around 8 tonnes of $CO_2$equivalent per year per person to 2 tonnes of $CO_2$equivalent per year per person. This transition requires structural changes, not merely the optimisation of existing systems.
- From an economic perspective, harmful effects on the environment, such as air pollution or the overexploitation of water resources, are not priced in the market and may therefore appear to be free, universal and infinitely available, unlike other goods and services. In other words, material well-being is currently valued more highly by society than natural resources. Thus, placing a legal, social or financial cost on the use of environmental resources could constitute a profound upheaval of current societal values and could be perceived as a new constraint.

[1] Research Director at Inria and mathematician at Sorbonne University, and also Deputy to the Scientific Director of Inria.
[2] Project Lead for 'International Governance and Foresight' within the AI Programme at Inria.
[3] Project leader for 'Digital Foresight and the Environment' within the Digital and Environment Programme at Inria.
[4] OECD (2019). *Council Recommendation on Artificial Intelligence*.

- While climate-related hazards are numerous and highly visible in France, the country benefits from a temperate climate and a developed economy, which enable adaptation. Such an approach is impossible in other parts of the world, which are affected by phenomena such as rising sea levels or extreme heat, rendering areas uninhabitable. The number of people directly affected seems small and they appear far removed; it is often through the media that the French come to realise the scale and consequences of disasters. This indirect impact tends to make awareness and good intentions regarding environmental conservation short-lived.

## Infrastructure inertia and the budgetary dilemma of adaptation

France aims to achieve carbon neutrality by 2050. In 2026, this ambitious target is medium-term on the scale of a human lifetime, but short-term on the scale of the evolution of much of the infrastructure (transport, housing, industry, energy). It is worth noting that the average age of a tractor in agriculture is 28 years, and that a nuclear power station will remain operational for around 50 years. This creates a certain degree of inertia in these sectors, which find it difficult to adapt to rapid changes, requiring long-term planning that takes into account shorter-term constraints.

In the face of climate change, science often proposes long-term solutions, which do not always address immediate environmental challenges. One examples consists in planting tree species in French forests, that are better suited to the future climate seems sensible, but will only produce significant effects after several decades. While the consequences of climate change are already very much upon us and require rapid responses, it is tempting to push these long-term measures into the background in favour of measures designed to react to climate change, such as compensation for those affected by disasters, geoengineering[5] and climate interventions. In other words, to focus on short- and medium-term solutions at the expense of long-term ones.

Whether it is a question of combating or adapting to climate change, the investments required across many sectors (mobility, housing, industry, energy, agriculture and food) are considerable. The costs arising from climate-related events are rising sharply, yet France is seeing its budgetary margins shrink, so that a dilemma is emerging: the rising cost of inaction on climate change is pitted against the colossal sum of investment required. An analogy would be an individual who knows that owning their own home will cost them far less in the long run than renting, but who, due to insufficient income, pays monthly rent.

Without glossing over the negative aspects of its development and use, the rest of this document seeks to show how digital technology and artificial intelligence can contribute to these transformations and to the ecological transition.

---

[5] A. Abecassis, P. Lamy (June 2026). *Geoengineering is a dangerous idea that we can no longer avoid*. Le Grand Continent, https://legrandcontinent.eu/fr/2026/06/23/geo-ingenierie-une-idee-dangereuse-que-nous-ne-pouvons-plus-eluder/

# 2. The environmental impacts of digital technology and AI: reality and misconceptions

## A growing environmental footprint, whose scale remains uncertain

In 2022, the France's Agency for Ecological Transition (ADEME) estimated that digital technology accounted for 4.4% of France's carbon footprint.[6] while this percentage is rising, it remains lower than that of other sectors such as transport (30%) or food (20%).

As regards the net environmental impacts of digital solutions for decarbonising other economic sectors, assessing these remains difficult, particularly due to their potential negative externalities. ADEME and its partners have quantified these impacts across an initial set of digital use cases: while net environmental gains are achieved in terms of carbon emissions, they remain modest compared with the carbon footprint of the economic sectors in question, and are sometimes associated with shifts in impacts, particularly regarding the depletion of abiotic resources.[7] ADEME notes, however, that these results cannot be generalised and that digital technology could be an ally in the ecological transition, provided that a prudent approach is adopted and potential negative effects are anticipated prior to the implementation of solutions.

The development of AI is currently playing a significant role in the growth of the digital footprint. On the one hand, the use of data centres increases water and electricity requirements; on the other hand, the supply of hardware components (graphics cards, accelerated replacement of devices, etc.) boosts demand for mineral resources. According to the International Energy Agency (IEA), data centres accounted for 1.5% of global electricity demand in 2025 and this figure could rise to 3% by 2030[8] .

The IEA emphasises that AI will drive faster innovation in the energy sector. This is expected to result from improved management and maintenance of electricity grids – which are becoming increasingly decentralised – as well as more accurate forecasting of generation capacity and consumption. However, it also highlights the limitations of the projections, particularly due to potential indirect effects that could offset the gains made, or barriers to adoption (restricted access to data, lack of infrastructure, societal obstacles, etc.).

It should be added that the hardware and software technologies used to develop AI systems, particularly foundational models, are constantly evolving and are accompanied by very significant performance gains (accelerators, quantisation, etc.). These not only reduce electricity consumption but also enable the development of increasingly large models.

Furthermore, the distinction between the environmental impacts of AI and those of digital technology in general might become increasingly blurred. Indeed, devices and technologies that make partial use of AI are now present in numerous digital tools (networks, softwares, applications, sensors, etc.) included in everyday objects. Focusing solely on the environmental performance of a few language models therefore provides a narrow view of AI development.

---

[6] Brilland, T., Fangeat, E., Meyer, J., & Wellhoff, M. (2025). *Assessment of the environmental impact of digital technology in France*. ADEME. https://librairie.ademe.fr/changement-climatique/7880-evaluation-de-l-impact-environnemental-du-numerique-en-france.html

[7] Camps, C., Delmas-Orgelet, J., Ekchajzer, D., Eskenazi, L., Fangeat, E., Fourboul, E., Le Goff, K., Lefevre, F., Malpart, G., & Roussilhe, G. (2025). *Direct and indirect environmental quantification of digital impacts in use cases – Digitalisation of nine use cases – Summary.* https://librairie.ademe.fr/economie-circulaire-et-dechets/9351-direct-and-indirect-environmental-quantification-of-digital-impacts-in-use-cases.html.

[8] International Energy Agency. *Artificial Intelligence.* Retrieved 8 September 2026, from https://www.iea.org/topics/artificial-intelligence.

## The need for transparency and environmental indicators

When it comes to the environmental impacts of the systems they develop, transparency is not a priority for many players in the AI sector, even though it would enable the phenomenon to be quantified, measured and addressed through improvements. In addition to electricity consumption, information such as how cooling systems operate, the frequency of training phases, the lifespan and the number of processors is rarely disclosed, forcing us to rely on incomplete estimates. Numerous comparisons are circulating to raise awareness amongst users of AI systems, such as those indicating the amount of water required for cooling for each query made on a generative AI system. While these figures have been obtained rigorously, they are valid only for a specific type of query on a specific system and cannot easily be generalised[9] . Indeed, the energy required to perform tasks can differ by several orders of magnitude between a simple text query and a query requiring image generation.

In order to quantify the environmental performance of AI systems and thereby enable their responsible use, greater transparency is required from all players in the digital sector. This would make it possible to ascertain the various forms of resource consumption (water, minerals, energy, etc.) throughout the life cycle of a given digital tool or service, i.e. from manufacture through to end-of-life, including the usage phase. To facilitate the responsible use of AI tools, but also to enable AI tool developers to adopt a resource-efficient approach, it would be useful to provide, for every query made or at every stage of development, quantitative information regarding the algorithmic complexity of the questions asked or the techniques employed.

While highly ambitious international regulation seems unrealistic at present, digital sector stakeholders adopting good environmental practices could usefully be encouraged. Environmental performance labels or indicators represent a promising avenue. Indeed, although the development of AI systems raises many concerns, transparency regarding the environmental impacts of these systems is essential, if only to encourage a coalition of voluntary stakeholders. In this spirit, the Coalition for Sustainable AI – launched in February 2025 by France, the United Nations Environment Programme and the International Telecommunication Union – provides documentation on its website, including guides to sustainable AI and a set of tools for measuring and reporting the environmental footprint of deployed systems[10] .

These challenges arise at a time when the need to live within planetary boundaries is becoming increasingly urgent.

---

[9] For further details, see: Luccioni, S., Trevelin, B., & Mitchell, M. (2024). *The environmental impacts of AI — Policy primer*. Hugging Face Blog. https://doi.org/10.57967/hf/3004

[10] Coalition for Sustainable AI. *International green AI initiatives*. Retrieved 8 September 2026, from: https://www.sustainableaicoalition.org/international-green-ai-initiatives/

# 3. Living in a constrained world

## The increasing scarcity of resources as a driver of a chosen frugality, for the benefit of society

The digital sector, as well as other sectors, has developed against a backdrop of abundant raw materials, energy and unhindered trade. While this relative abundance has obviously not encouraged a frugal lifestyle, the situation could change:

- Demand for certain metals is rising sharply, yet supply is limited (due to the depletion of reserves and export quotas imposed by a growing number of countries)[11] , and the opening of new mines is facing public resistance over their environmental impact ;
- Permanent magnets made from rare earths are now essential components of aerospace technologies, digital technologies (mobile phones, hard drives, etc.) and technologies linked to the energy transition (electric motors, generators for wind turbines, etc.). According to the IEA, in 2024, China refined 91% of the rare earths used in permanent magnets and produced 94% of these magnets[12] . This quasi-monopoly thus gives China significant leverage over the pace of technological development in other countries;
- Energy prices and incentives to use renewable energy could increase;
- The regulatory landscape could change ; as the impacts of climate change become more pronounced, stricter enforcement of existing measures and regulations is a possibility[13] .

These trends suggest a shift towards a world of constraints, where resource efficiency will certainly become a constraint, but also an asset, both in terms of competitiveness and strengthening sovereignty. An eco-responsible digital sector, one that uses raw materials and energy efficiently, will therefore make it possible to combine economic and environmental performance, thus contributing to the well-being of present and future generations. Developing an eco-responsible digital sector is a fundamental investment in the future for the whole industry that designs and manufactures digital devices and develops software and applications.

*Low-tech*, which can be defined as digital technology that is appropriate, frugal and resilient, has a bright future ahead. Nowadays, the race for performance – faster microprocessors, feature-rich and resource-intensive software, and telecommunications networks carrying ever-increasing volumes of data – is accelerating the replacement cycle of digital equipment[14] . Yet many of the everyday tasks performed by a computer or a mobile phone require neither large volumes of data nor high processing speeds, and do not call for sophisticated features. *'Low-tech'* does not refer to second-rate technology, but is, on the contrary, a major area of research for the scientific community, calling for a holistic approach. To be sustainable, all digital components – from end devices to software, including networks and sensors – must be *low-tech*.

---

[11] In its baseline scenario for 2035, the IEA forecasts a supply shortfall relative to demand of 32% for lithium, 26% for cobalt and 25% for copper.
See: International Energy Agency (2026). *Global Critical Minerals Outlook 2026*. IEA. https://www.iea.org/reports/global-critical-minerals-outlook-2026.

[12] International Energy Agency (2026). *Rare Earth Elements*. IEA. https://www.iea.org/reports/rare-earth-elements

[13] It is worth recalling the constitutional significance of the 2004 Environmental Charter, which, amongst other things, establishes a duty to 'contribute to the preservation and improvement of the environment' (Article 2).

[14] It should be noted that 50% of the digital sector's carbon footprint, in France, is linked to the manufacture and operation of devices (televisions, computers, smartphones, etc.), 46% to data centres and 4% to networks; see: Brilland, T., Fangeat, E., Meyer, J., & Wellhoff, M. (2025). *Assessment of the environmental impact of digital technology in France*. ADEME. https://librairie.ademe.fr/changement-climatique/7880-evaluation-de-l-impact-environnemental-du-numerique-en-france.html

It is important to keep in mind the 'total' scope of AI, which calls for an interdisciplinary approach: its development trajectory is not determined solely by the results of scientific research and technological advances. These must be considered alongside human and social perspectives, as consumers' adoption of technologies and the ways they use them largely shape the developments observed. The work of economists, sociologists, psychologists and legal experts must therefore help us analyse our relationship with digital technology and understand the transformations it brings about in our societies and in the way we relate to the world. In doing so, it will enable public policy-makers to identify levers for action to anticipate the consequences of these technologies and to ensure that rights and freedoms are respected.

## Specialisation and restraint in usage as the keys to frugal AI

The race towards ever-greater scale is not inevitable, and its effects should be compared with those of specialised models, whose size is tailored to their intended use. while AI is not a new technology, the development of large generative models is, and the launch of ChatGPT in November 2022 sparked a race for for sheer scale: the number of parameters, the volume of data used, the number of queries processed, and so on. The emergence of a new technology with numerous potential applications, such as AI, initially encourages the design of systems capable of meeting a wide range of expectations. Subsequently, constraints relating to performance and accuracy, along with the desire for systems that are easy to develop and simple to use, lead to a shift towards specialised systems capable of efficiently performing a limited number of tasks. This trend appears to be evident at present with the development of AI systems tailored to specific application domains (law, geography, agriculture, etc.). Specialisation reduces the size of AI models developed in this way – and thus their environmental impact – while making them easier to build and manage. This development also fosters innovation, by enabling many new players to emerge and offer dedicated solutions, whereas an oligopoly of large AI models owned by a few players would stifle innovation. Finally, an AI system built on a limited volume of data whose quality, provenance and ownership are guaranteed, also provides security, particularly from a legal perspective.

Given the financial resources and computing power possessed by certain digital giants, choosing to develop frugal AI technologies is a promising strategy, which requires the development of alternative approaches. Being resourceful and doing more with less are motivating scientific challenges, especially for young scientists.

Managing environmental impacts also relies on training, the development of educational content and the creation of labels, to inform users' choices and purchases.[15] The ease of use of many generative AI tools has led us to forget that existing information – such as a recipe or a poem by Victor Hugo – does not need to be generated, as it is already available on numerous websites. There is no need to try to learn every recipe or every poem by Victor Hugo, when all you need to do is find out where they are available.

Jean Giono wrote, 'The embankment that runs alongside my road is richer than Oceania'. It is not a question of curbing our curiosity or our interactions, but rather of changing the values that determine our choices: is something that is new, disposable, powerful, energy-intensive and produced far away, necessarily better?

[15] A MOOC (in french) is available to help you familiarise yourself with the environmental impacts of digital technology: https://www.fun-mooc.fr/fr/cours/impacts-environnementaux-du-numerique/.

Indeed, AI also presents an opportunity in science and knowledge that can be harnessed to support the ecological transition.

# 4. AI as a catalyst for scientific discovery

## A new way of interacting with data to drive scientific progress

Whether it concerns meteorology, climate change and its consequences, or natural and environmental risks, digital technology plays a major role in understanding and predicting the phenomena around us. Through the collection of diverse and precise measurements, the construction of models and their numerical simulation, or the analysis of available data, digital technology is essential to our understanding of geophysical phenomena. Without it, what would we know about the consequences of human activities on the environment? What policies could we develop to address them? What would an IPCC report be without digital technology?

In many fields, AI enables major scientific advances which do not merely build on existing work, but also offer a new way of conceptualising the studied objects.

For centuries, the development of scientific knowledge has relied on the observation of characteristic phenomena, followed by the construction of mechanical, physical and mathematical models to represent them, in order to compensate for the limited amount of observational data available. Archimedes' principle, Galileo's law and Newton's theory of gravity are classic examples of this approach, which also underpin the formalisation of knowledge and its teaching. While the earliest models were relatively simple, the complexity of nature and of the phenomena around us has led to increasingly sophisticated models that are difficult to handle, parametrise and study[16]. The development of increasingly sophisticated models is hampered by the difficulty of validating them and extracting relevant information from them.

Since around the beginning of the 20th century, advances in metrology and the availability of vast amounts of digital data (from sensors, satellites and digital traces left on the internet) have drastically altered this landscape. With accurate, abundant and varied data, it is possible to characterise and track changes in the systems under study. AI, and more generally data science, enables us to extract from all these available observations meaningful information that characterises the phenomena and allows us to predict their evolution. In the field of meteorology, forecasting models based on data collected over decades have been proposed[17] and produce high-quality results.

In some cases, data-driven modelling supersedes the traditional approach, but a hybridisation of the two approaches is often observed: the mechanistic model provides the low-frequency trends, while data-driven modelling provides the high-frequency component of the phenomena. If man-made structures are relatively easy to model mathematically (a bridge, a car, etc.), the traditional modelling approach has shown its limitations when it comes to representing the living world (ecosystems, collective dynamics, etc.), and the role of data-driven models is set to grow, opening up new avenues and providing fresh impetus.

[16] One example relates to fluid mechanics, where taking into account extended physical properties makes it difficult to analyse and solve established theoretical models.
[17] For example, the GenCast model, developed by Google DeepMind, or the ArchesWeather model, developed by researchers at Inria.

## AI can be a tool for the ecological transition

AI enables major scientific advances and opens up new avenues in numerous fields, all of which share the common feature of involving phenomena difficult to model as they are characterised by a large number of parameters, complex mathematical modelling, variability in individual behaviour, and the presence of living organisms. As such, these fields are ideal areas for the development of AI systems:

- Renewable energy: optimisation of production, demand forecasting, electricity grid balancing, predictive maintenance;
- Agriculture: development of decision-support tools; genetic diversity to help species adapt to climate change; robots performing arduous tasks;
- Consultancy and expertise: development of language models which, based on a corpus of texts and data, will provide detailed analyses, identify emerging trends and answer complex questions;
- Home automation: smart energy management, early detection of anomalies, enhanced comfort levels (air quality, humidity, temperature);
- Biodiversity: monitoring the functioning of ecosystems and their interactions, predicting changes in biodiversity, and developing indicators.
- Health: new molecules, new treatments, assisted surgical procedures, remote patient monitoring, smart prosthetics, personalised treatments…

# 5. The transformations brought about by AI

Digital technology has achieved the remarkable feat of making highly sophisticated tools, that are capable of complex tasks, easy to use. The power of digital tools, and the close relationship we have with them through the personal data we entrust to them, make them effective levers for changing our habits, behaviours and values, possibly in favour of environmental protection.

## A driver of knowledge dissemination and access

There is an entire area of science that has been transformed by AI : the dissemination- and interaction with knowledge, which is profoundly changing the way knowledge is transmitted and decisions made, particularly in environmental matters.

Researchers, businesses and other organisations produce a rich and varied body of scientific literature: scientific publications, technical reports, data and more. Despite the availability of digital sharing platforms[18], this literature is often confined to a small circle of specialists. Only the few topics that attract media attention reach the public debate, leaving a large proportion of scientific advances in the shadows. Several factors explain this limited dissemination of knowledge:

- Complexity and technical nature: publications, often in English, are difficult to access for the general public;
- Fragmentation of results: each study provides a partial, incremental answer to a problem ;
- Economic barriers: access to certain articles in prestigious journals is subject to a fee;

[18] Similar to data.gouv.fr for data or HAL (https://hal.science/) for scientific publications.

- Volume of publications : around 2.8 million articles were published in 2022, with varying levels of quality and originality, making the information difficult to grasp.

Digital tools can be powerful agents of change, particularly through recommendation algorithms, which are largely based on AI. Faced with the accumulation of knowledge, its complexities and interdependencies, citizens, professionals and decision-makers need to inform their choices with scientific evidence. This need is particularly evident in areas crucial to our well-being, such as food, transport, agriculture, the environment and health. While written documents remain the primary means of formalising and disseminating scientific knowledge, AI and chatbots are opening up unprecedented possibilities for the dissemination of knowledge:

- Compilation and synthesis: providing information that is relevant and tailored to the enquirer's level of technical expertise;
- Natural language interaction: simple queries and dialogue to refine responses.

While the development of these conversational agents raises numerous questions – particularly regarding bias, hallucinations, intellectual property and liability – the quality of the answers provided heralds a new era in our interaction with knowledge. These tools have become the primary gateway to knowledge.

AI is bringing about a new paradigm shift that extends beyond the scientific sphere and permeates society as a whole, thereby presenting an opportunity to shape societal choices. For activities that have been practised for a long time and are the result of a long evolution, innovation often takes the form of a new development that is integrated into, or improves upon, a sequence of tasks. But as J. Schumpeter described, innovation can also possess a force of 'creative destruction' that opens up a new path competing with the old one, and sometimes destroying it. Gas light and the electric light bulb, which supplanted the candle, are examples of this. Digital technology often enables such disruptive changes, such as search engines revolutionising commerce or geolocation transforming the way people travel.

## Automation as a driver of transformation in decision-making

While AI can be a powerful tool for disseminating knowledge and informing decision-making, it presents another major disruption: it is the first time that humans have designed a digital system over which they may lose control. Furthermore, it is also driving a shift from a 'Cartesian' mode of reasoning (hypothesis, logic, argumentation) to a pragmatic and statistical approach. A statistical logic and an approach based on extrapolation from big data risk marginalising long-term planning, acts in favour of short-term decision-making.

Agentic AI models[19] , which have emerged recently, go a step further by shifting from responding to requests to acting autonomously. These systems are capable of perceiving their environment, planning a sequence of actions and executing them by calling upon external tools (web browsing, APIs, etc.). Removing humans from the decision-making loop makes it more difficult to oversee these non-deterministic systems and to reconstruct the decisions taken.

Given that lines of reasoning and decision-making are evolving, it is essential that the AI systems underpinning this evolution incorporate the objectives of the ecological transition right from the design stage.

---

[19] Official report of the French G7 Presidency, coordinated by Inria (June 2026). *Agentic AI: Deployment, Adoption and Impacts*. https://www.entreprises.gouv.fr/files/files/Actualites/2026/g7/agentic-ai-inria-short.pdf

## Changes to the value chain of work and reasoning

In the realm of intellectual reasoning, what role is there for humans in the face of AI?

In mathematics and philosophy, where abstract reasoning plays a significant role, AI systems no longer merely provide assistance with writing or literature reviews, but produce powerful and verified results. Indeed, AI is capable of statistical reasoning, based on large volumes of data, as well as constructing proofs for conjectures or theorems. Thus, by 2026, AI had produced numerous proofs of mathematical problems that had remained open for decades, such as the existence and regularity of solutions to the Navier–Stokes equations[20]. On the one hand, future proofs of major scientific results could be obtained by AI and on the other, a consequence of AI is also a disruption of the value chain. Beyond the field of scientific research, this development is affecting many professions in which tasks involving analysis, synthesis, data collection and consultancy are carried out by AI systems. AI is transforming these affected fields from a form of craftsmanship to industrialisation, bringing about a profound change in production processes and objectives.

What value should we assign to what AI does quickly, easily and automatically? The production of a limited number of results – which are difficult to obtain and each of which has significant value – risks being replaced by the production of a vast number of results, arising from a process that is simple to implement and therefore each of which has reduced value. This reversal of values presents an opportunity for profound changes in our societies.

The 19th century and the Industrial Revolution saw the shift of workers from the primary sector (agriculture) to the secondary sector (industry). During the 20th century, the number of jobs in the tertiary sector (services) grew significantly, at the expense of the other two sectors. As further changes to the labour market are underway, the question arises as to the emergence of a quaternary sector, forcing us to rethink the place of work in society. While technological developments have gradually led to a dehumanisation, both in the workplace (the use of computers and software) and in everyday life (self-service terminals and checkouts, online shopping, etc.), AI cannot be held solely responsible for the loss of meaning in many jobs and tasks. It has simply taken a long-standing process a step further. In the face of these inevitable developments, the re-humanisation of certain professions (retail, services) and the skilled trades – including for intellectual tasks – appears to be an appropriate response. As early as 1997, Roger Sue proposed a 'quaternary sector' in which human relationships would be valued more highly than goods[21] . A reorganisation of our system along these lines could also place greater value on professions linked to environmental protection.

**The developments described above are all-encompassing; they call into question the structures and ways of operating to which we are accustomed, and to which we have contributed. The scientific and technical progress made over decades and the advent of AI give us a glimpse of the end of a cycle. The next cycle is largely yet to be shaped, and now is the right time to propose new directions, a different trajectory and alternative paradigms. If 'our legacy is not preceded by any will' (René Char), let us seize the opportunities presented by AI to chart a desired course of development.**

---

[20] OpenAI (2026). *On the Navier-Stokes Millennium Prize Problem.* https://openai.com/fr-FR/index/navier-stokes-solution/.

[21] Sue, R. (1997). *The Wealth of Mankind: Towards the Quaternary Economy*. Odile Jacob. https://shs.cairn.info/la-richesse-des-hommes--9782738105202?lang=fr .

*This text was originally written in French without the aid of artificial intelligence. However, DeepL has been used to help with translation. Perhaps this is a mistake; it might have been better structured and more convincing had it been generated or revised by an AI tool. This contribution sets out an analysis, considers possible developments and proposes directions to promote the ecological transition, but it offers no specific formula and no quantitative assessment.*